\documentclass[12pt,preprint2]{aastex}

\usepackage{amssymb}
\usepackage{epstopdf}
\usepackage{url}

\shorttitle{Galaxy query by example}
\shortauthors{Lior Shamir}

\begin{document}

\title{Morphology-based query for galaxy image databases}
 
\author{Lior Shamir\altaffilmark{1}}
\affil{Lawrence Technological University, Southfield, MI, 48075}
\email{lshamir@mtu.edu}

\altaffiltext{1}{Lawrence Technological University, Southfield, MI, 48075}


\begin{abstract}
Galaxies of rare morphology are of paramount scientific interest, as they carry important information about the past, present, and future universe. Once a rare galaxy is identified, studying it more effectively requires a set of galaxies of similar morphology, allowing generalization and statistical analysis that cannot be done when $N=1$. Databases generated by digital sky surveys can contain a very large number of galaxy images, and therefore once a rare galaxy of interest is identified it is possible that more instances of the same morphology are also present in the database. However, when a researcher identifies a certain galaxy of rare morphology in the database, it is virtually impossible to mine the database manually in the search for galaxies of similar morphology. Here we propose a computer method that can automatically search databases of galaxy images and identify galaxies that are morphologically similar to a certain user-defined query galaxy. That is, the researcher provides an image of a galaxy of interest, and the pattern recognition system automatically returns a list of galaxies that are visually similar to the target galaxy. The algorithm uses a comprehensive set of descriptors, allowing it to support different types of galaxies, and it is not limited to a finite set of known morphologies. While the list of returned galaxies is neither clean nor complete, it contains a far higher frequency of galaxies of the morphology of interest, providing a substantial reduction of the data. Such algorithms can be integrated into data management systems of autonomous digital sky surveys such as the Large Synoptic Survey Telescope (LSST), where the number of galaxies in the database is extremely large. The source code of the method is available at \url{http://vfacstaff.ltu.edu/lshamir/downloads/udat}.
\end{abstract}




\section{Introduction}
\label{introduction}

Galaxy morphology is critical for understanding galaxy evolution, galaxy interactions, and studying new forms of extragalactic objects. The ability to collect large databases of galaxy information has enabled the studying of some of the most fundamental questions about the universe such as profiling its large scale structure and characterizing its physical properties \citep{colless20012df}. 

Robotic telescopes can acquire and store very large astronomical databases, and the size of these databases is expected to grow further when powerful imaging devices such as LSST see first light. Future space-based missions with wide-angle field of view such as the Wide-Field Infrared Survey Telescope \citep{spergel2013wide} are also expected to generate large astronomical databases. A substantial portion of these data is in the form of images, reinforcing  the need for developing automatic image analysis methodology that can process these large databases and turn them into scientific discoveries \citep{edwards2014astronomy}.

One approach to analyzing very large databases of galaxy images is by utilizing the analysis power of human volunteers who access the data via a web-based interface to produce catalogs of manually annotated galaxies \citep{lintott2011galaxy,keel2013galaxy,willett2013galaxy}. However, the bandwidth of manual annotation cannot satisfy the data collection capacity of the current digital sky surveys such as the Dark Energy Survey (DES), and its reliance on the processing power of the human brain limits the opportunities to improve its bandwidth, making it even more difficult to effectively analyze future sky surveys such as LSST. That reinforces the use of automation to generate catalogs of galaxy morphology \citep{shamir2009automatic,dieleman2015rotation}, and automatically generated catalogs have been collected and published \citep{hue10,fasano2012morphology,shamir2014automatic,gravet2015catalog,kuminski2016computer,huertas2016mass}.

However, automatic annotation of galaxies by their morphology is merely one task related to galaxy image analysis that can be performed by computers. Other tasks can include automatic detection of peculiar galaxies in large datasets \citep{shamir2012automatic,shamir2014automatic}, or grouping galaxies by their visual similarities using unsupervised machine learning \citep{shamir2013automatic,schutter2015galaxy}.

One of the tasks that can be extremely difficult to perform manually is searching galaxy image databases for peculiar galaxies of a certain morphology of interest. For instance, studying a system such as Arp 142 \citep{arp1987catalogue} can be more productive if the researcher has a set of morphologically similar systems, so that she can compare and identify patterns or measurements that are typical to that specific system and distinguish it from other systems. However, identifying a set of systems that are visually similar to Arp 142 in databases of millions or even billions of galaxies is virtually impossible to perform without automation. Clearly, peculiar galaxies are not necessarily interacting systems, and examples are polar ring galaxies \citep{whitmore1990new}, dust-lane ellipticals \citep{sadler1985common,bertola1985warped}, or the peculiar ``Hanny’s Voorwerp'' \citep{lintott2009galaxy}.

\begin{figure}[ht]
\begin{center}
\includegraphics[scale=1.00]{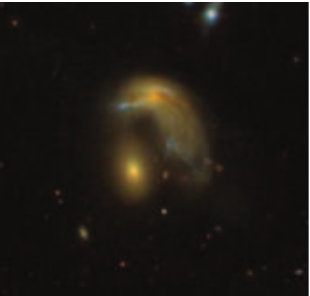}
\caption{The Arp 142 system imaged by Sloan Digital Sky Survey}
\label{arp142}
\end{center}
\end{figure}

\begin{figure}[ht]
\begin{center}
\includegraphics[scale=1.00]{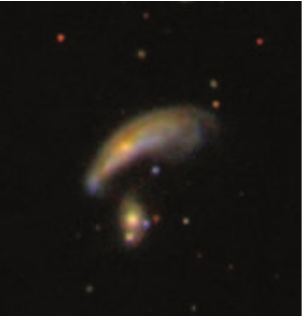}
\caption{A system visually similar to Arp 142}
\label{not_arp142}
\end{center}
\end{figure}

A potential automatic method that allows better studying of a system such as Arp 142 can take an image of the system as shown in Figure~\ref{arp142} as input, and return a list of interacting systems that are visually similar to it. Such automatic system can return a list of interacting systems such as the one displayed in Figure~\ref{not_arp142}, found automatically by an algorithm that mined for peculiar galaxy pairs in SDSS \citep{shamir2014automatic}. A list of such interacting systems could allow the studying of systems such as Arp 142 more effectively, and with $N>1$. Such algorithms do not necessarily require completeness, as in very large databases it can be assumed that even very rare objects will occur multiple times, and therefore even finding a fraction of these instances can provide sufficient data to study a specific rare system.

While some peculiar systems of interest are very rare, with the power of robotic telescopes such as LSST even a rare one-in-a-million type of galaxy would occur $\sim$10,000 times in its database of $\sim$10 billion galaxies. It is therefore clear that the information required to study these systems is contained in the database, but needs to be detected. Here we describe an algorithm that takes a target galaxy image as input, and mines through datasets of galaxy images to return a list of galaxies that are visually similar to the target galaxy.

\section{Image analysis method}
\label{method}

The automatic identification of galaxy images that are visually similar to a certain query galaxy is performed by computing the dissimilarity between the query galaxy and each of the other galaxies in the database. The measured dissimilarity values between the query galaxy and all galaxies in the database are then sorted to return a list of the galaxies with the smallest computed dissimilarity, and are therefore assumed to be the most similar to the query. Since the query galaxy is not known when the system is trained, supervised machine learning methods such as support vector machine (SVM) or deep learning do not provide a natural solution that can address the identification of galaxies with morphologies not known at the time of training.


\subsection{Numerical image content descriptors}
\label{features}

The first step in the detection process is the conversion of each galaxy image into a set of 2881 numerical image content descriptors computed using the Wndchrm feature set \citep{Sha08}. The Wndchrm feature set is a mature comprehensive set of numerical image content descriptors that includes various numerical characteristics of the visual content. It includes textures, polynomial decomposition of the pixel intensities, statistics of the pixel intensities, fractals, high-contrast features, and more, as thoroughly described in \citep{Sha08,orlov2008wnd,shamir2010impressionism,shamir2013automatic,shamir2012automatic,shamir2012computer}. That scheme provides a comprehensive numerical reflection of the visual content, and has been found effective for several tasks related to automatic analysis of galaxy morphology such as automatic annotation \citep{shamir2009automatic,kum14}, unsupervised analysis of galaxy morphology \citep{shamir2013automatic,schutter2015galaxy}, and peculiar galaxy detection \citep{shamir2012automatic}. These methods were also applied to produce  catalogs of galaxy morphology \citep{shamir2014automatic,kuminski2016computer}.

The Wndchrm feature set also contains color descriptors \citep{shamir2012art,shamir2012computer,shamir2010impressionism}. However, these color features have shown mild contribution to the task of galaxy image analysis \citep{shamir2009automatic,kum14,shamir2014automatic}, and are therefore not used in this experiment. All images are treated as grayscale images.

\subsection{Dissimilarity measurement}
\label{distance}

After each galaxy image is represented by a vector of numerical values that reflect its visual content, the distance between the feature vector of the target galaxy image and the feature vector of each of the galaxies in the dataset is computed. That allows selecting the galaxies with the shortest distance to the feature vector of the target galaxies. Two different methods for measuring the distance in a multi-dimensional space were used:\newline \newline
1. Weighted Euclidean Distance: The  Weighted Euclidean Distance $d$ between feature vector $X$ and feature vector $Y$ is a simple measure defined by $d=W\sqrt{\Sigma_i (X_i-Y_i)^2}$, where $W$ is a vector of feature weights that reflects the informativeness of each feature as will be described later in this section. \newline
2. Earth Movers Distance (EMD): The Euclidean distance is based on the assumption that each feature in the feature vector is an independent measurement. However, most of these features are histogram bins \citep{Sha08}, and therefore important information in the feature vectors is not used when treating each value as an independent measurement.

EMD was found efficient for dissimilarity measures in image analysis using pattern recognition and multimedia retrieval \citep{rubner2000earth,ruzon2001edge}. It can be conceptualized as the minimum amount of work required to fill a distribution of holes in space with the mass of Earth distributed in the same space, such that a unit of work is the work required to complete the movement of an Earth unit by a distance unit. The problem can be formalized by Equation~\ref{emd}

\begin{equation}
\label{emd}
Work(X,Y,F)=\Sigma_{i=1}^n \Sigma_{j=1}^n f_{i,j}d_{i,j},
\end{equation}

where X and Y are the weighted feature vectors ${(Wx_1,x_1).....(Wx_n,x_n)}$ of size n, and $f_{i,j}$ is the flow between $X_i$ and $Y_j$. The flow F can be determined by solving a linear programming problem with the following constraints: \newline \newline
$Wx_i \geq \Sigma_{j=1}^n f_{i,j}$      \newline \newline
$Wy_j \geq \Sigma_{i=1}^n f_{i,j}$     \newline \newline
 $\Sigma_{i=1}^n \Sigma_{j=1}^n f_{i,j} = \min ( \Sigma_{i=1}^n Wx_i , \Sigma_{j=1}^n Wy_j ) $     \newline


The earth mover’s distance between X and Y is then defined as: \newline
$EMD(X,Y)=\frac{Work(X,Y,F)}{\Sigma_{i=1}^n \Sigma_{j=1}^n f_{i,j} } $

More information about EMD can be found in \citep{rubner2000earth,ruzon2001edge}. 

Most of the values in the Wndchrm feature vector are histogram bins \citep{Sha08}. The Wndchrm feature vector combines several different histograms into one vector. For instance, the Zernike features contribute 72 features to the vector, and the Chebyshev statistics features contributes a histogram of 32 features \citep{Sha08}. Therefore, measuring the dissimilarity between a pair of vectors using a single EMD comparison of the two full vectors might not provide an optimal similarity measure due to the extra work needed to equalize unrelated histograms. Moreover, many of the features in the feature vector are discrete, and are not histogram bins (e.g., the Tamura texture directionality), and dissimilarities between these features should not be measured using EMD as they have no link to the other features. 

To solve these two problems, we use a two-layer scheme of vector similarity measure, such that each pair of histograms (one from each feature vector) is measured using EMD. That is, the Zernike features of vector X is compared to the Zernike feature of vector Y using EMD, the Chebyshev features of vector X is compared to the Chebyshev feature of vector Y using EMD, and so on. All variables that are not histogram bins are compared using the weighted Euclidean distance. Then, the sum of all distances provides the measured dissimilarity between the two vectors. The EMD dissimilarity and Euclidean distance dissimilarity are two different dissimilarity measures, but since all features are weighted using the same weighting mechanism, the Euclidean distance and EMD dissimilarity can be combined into a single dissimilarity score.

\subsection{Feature weights}
\label{weights}

As described in Section~\ref{features}, the set of numerical image content descriptors that reflect the visual content is large and comprehensive, and therefore it is reasonable to expect that not all of these descriptors are equally informative, and some of them can be considered noise. As mentioned in Section~\ref{distance}, each feature is assigned with a weight that reflects its informativeness and determines its impact on the results. These weights are computed in this study in two different ways: \newline \newline

1. Variance: The variance is a crude heuristics of the feature weights, but has demonstrated some good results in tasks related to unsupervised machine learning with the large Wndchrm feature set \citep{shamir2012automatic,manning2014chloe}. The intuition of using the $\frac{1}{\sigma^2}$ as weights is that a feature with high variance is likely to be noisier than a feature with lower variance \citep{shamir2012automatic,manning2014chloe}. In case of a noisy feature with low variance, that feature will be assigned with a high weight, but because the variance is low it is more likely that the differences between the values computed for the target sample and the database samples will be lower, so the impact of such noisy features with low variance on the dissimilarity measure will be relatively small.

2. Entropy: The entropy weight $E_f$ of feature {\it f} is computed by \newline $E_f=1-\Sigma_i p_i \log_2p_i$, \newline where $p_i$ is the probability of a feature value to fall into bin {\it i}. The intuition of using the entropy is similar to the intuition of using the variance, but the entropy does not assume normal distribution of the feature values. That can lead to more efficient weights, as a database of random galaxies is expected to have different morphological types, but since the morphology of most galaxies is consistent \citep{hubble36,san61} most of these galaxies fall into a finite number of defined classes. Therefore, a feature that changes based on the morphology of the galaxy is expected to have a higher $E_f$.

\section{Data and performance evaluation}
\label{data}

The performance of the method can be measured by using two classes of galaxies; one is the query class and the other is the database class. The database class is the class of ``regular'' galaxies in the database, and the query class is the class of galaxies that the algorithm attempts to identify based on a query galaxy. The evaluation process is performed by combining $M$ galaxy images from the query class with the $N$ database galaxies. The algorithm can then be applied by selecting one of the $M$ query galaxies as the query galaxy, and combining a subset of the remaining $M$-1 query galaxies with the $N$ database galaxies. That process can be repeated up to $M$ times such that in each run a different galaxy is used as the query galaxy.

The hit rate performance evaluation is determined by Equation~\ref{performance}

\begin{equation}
\label{performance}
\frac{\Sigma_{m=1}^{|M|} \Sigma_{r=1}^{|R_m|} (R_{m_r} \in M \wedge R_{m_r} \neq m)} {|M|} ,
\end{equation}
such that M is the set of query galaxies, and $R_m$ is the set of galaxies returned by the algorithm as the most similar to query galaxy $m$. That is, any  galaxy of the query class in the top $R$ galaxies returned for a certain query galaxy is considered a hit. The hit rate is measured by the average number of galaxies of the query class in the list of top $R$ galaxies returned by the algorithm in each of the $|M|$ queries. That process is repeated sequentially such that in each run a different image $m$ is used as the query image, and the performance is measured by averaging the number of galaxies form the query class returned among the top R galaxies. 

The size of the returned list $R$ is the {\it rank}, and can be set to a different value in each experiment. 
The proposed system is not expected to be fully accurate, so that $R$ can also include galaxies that are not of the same morphological type as the query galaxy. However, the purpose of the method is to reduce the data such that the frequency of galaxies that are morphologically similar to the query galaxy is much higher in R than in the entire galaxy population in the database.

The data used for testing the system are galaxy images taken from Sloan Digital Sky Survey (SDSS), and downloaded automatically through the Catalog Archive Server (CAS) as described in \citep{kuminski2016computer}. The images are 120$\times$120 pixel JPEG images converted to the Tagged Image File (TIF) format \citep{kuminski2016computer}.

Several datasets were used in the experiments. The first is galaxies annotated automatically as spiral or elliptical galaxies \citep{kuminski2016computer}. In a universe with only early-type galaxies, a spiral galaxy can be considered ``peculiar'', so that the dataset can be used such that a small set of spiral galaxies are combined with a larger set of elliptical galaxies (or vice versa), and then a single spiral galaxy is used as the query image.

\begin{figure}[ht]
\begin{center}
\includegraphics[scale=0.55]{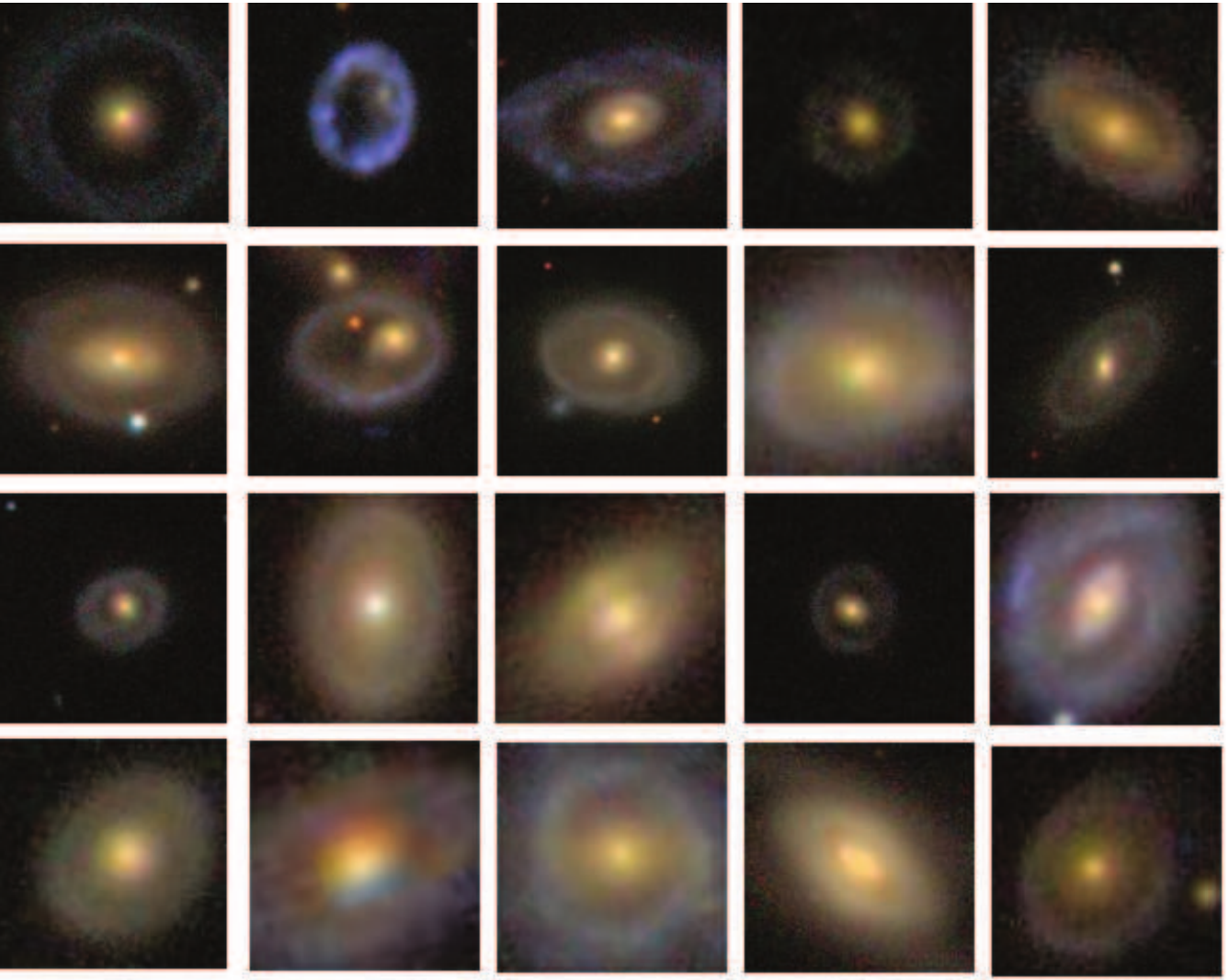}
\caption{The dataset of 20 ring galaxies}
\label{ring_galaxies}
\end{center}
\end{figure}

\begin{figure}[ht]
\begin{center}
\includegraphics[scale=0.55]{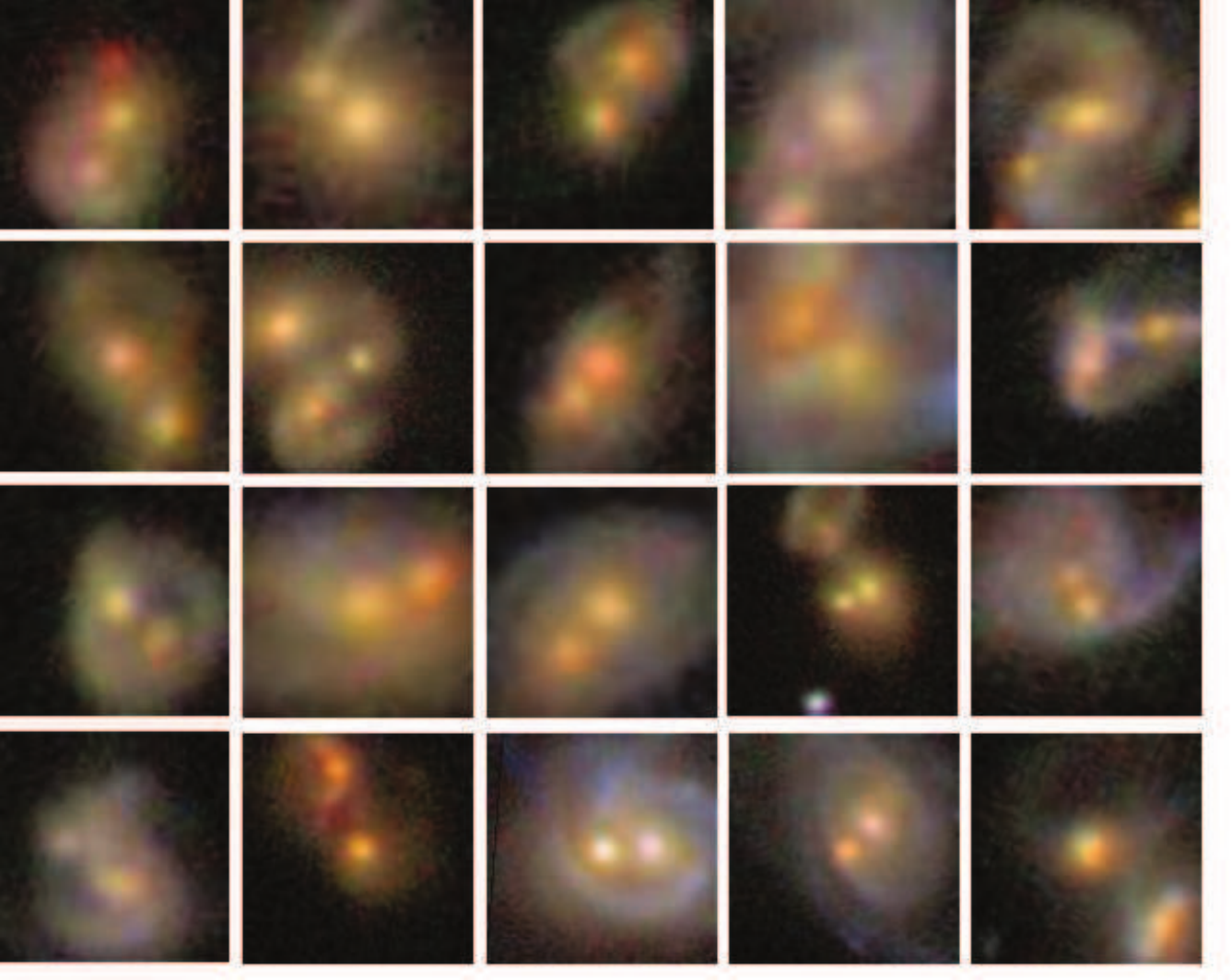}
\caption{Images of the 20 interacting galaxies in the dataset}
\label{mergers}
\end{center}
\end{figure}

A small dataset contained 100 spiral galaxies and 100 elliptical galaxies taken from \citep{kuminski2016computer}, and visually inspected to ensure that the two classes are consistent. These datasets were also used in combination with two smaller datasets of 20 ring galaxies and 20 interacting galaxies \citep{shamir2014automatic}. Figures~\ref{ring_galaxies} and~\ref{mergers} show the images of the 20 ring galaxies and the 20 interacting galaxies, respectively.

Additionally, a dataset of 4,000 galaxies classified as spiral and 4,000 galaxies classified as elliptical were used to test whether the ring and merger galaxies can be detected in a larger set of several thousand galaxies. The advantage of using the galaxies of \citep{kuminski2016computer} is that the galaxies are annotated, so that they can be used such that all galaxies in the database class are of the same broad morphological type (elliptical or spiral) and all galaxies in the query class are of the other broad morphological type. 

\begin{figure}[ht]
\begin{center}
\includegraphics[scale=0.6]{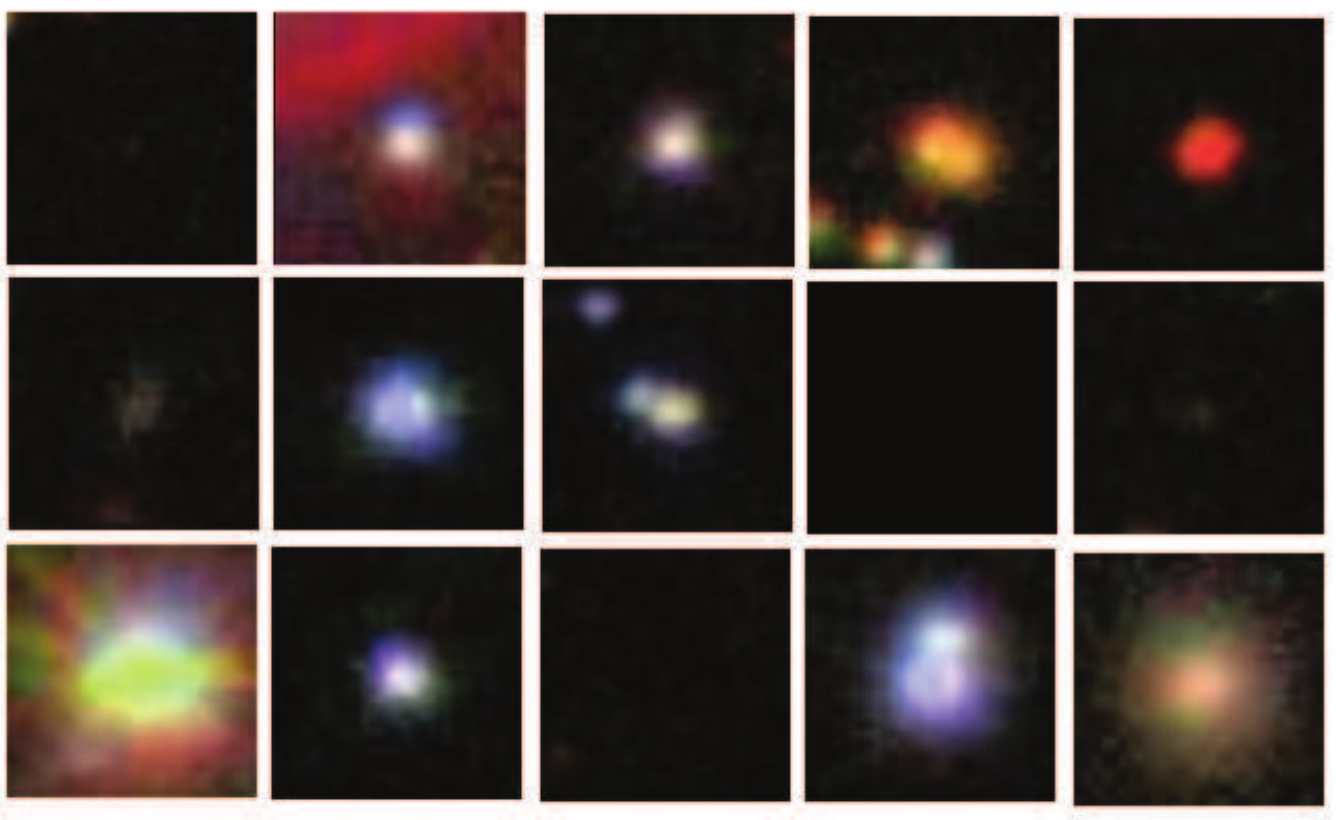}
\caption{The first 15 galaxies in the dataset of 10,000 galaxies with SDSS i magnitude less then 18.}
\label{i18_galaxies}
\end{center}
\end{figure}

Another dataset that was used contained 10,000 random objects classified by SDSS photometric pipeline as galaxies. These galaxies are not annotated in any way, and are therefore less consistent, providing a more diverse sample when used as the database class. On the other hand, a galaxy returned by the list as one of the R most similar galaxies but is not in the query set $M$ is not necessarily noise, because the database class is diverse and uncontrolled, and therefore can contain also galaxies that happen to be similar to the query galaxy. Since the vast majority of SDSS galaxies are small and faint, the dataset contained just galaxies with SDSS i magnitude brighter than 18, so that the algorithm is not able to perform well by simply comparing the brightness of the objects. Figure~\ref{i18_galaxies} displays the first 15 galaxies (when ordered by the galaxy ID) in the dataset, showing the diversity of the objects included in that dataset.

\section{Results}
\label{results}

\begin{figure}[ht]
\begin{center}
\includegraphics[scale=0.55]{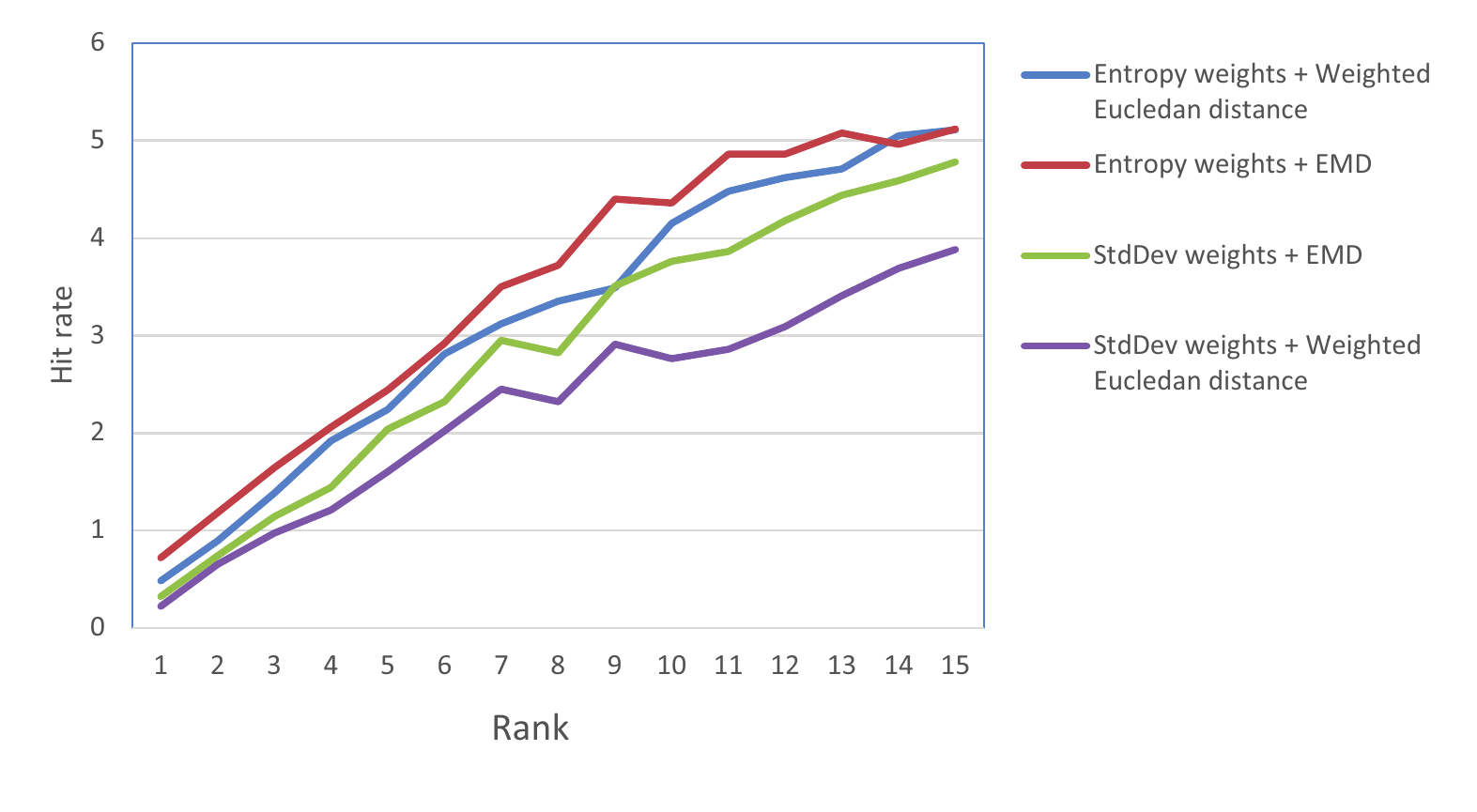}
\caption{Hit rate when using 100 elliptical galaxies as ``regular'' galaxies and 10 spiral galaxies as ``peculiar'' galaxies. The hit rate is determined by the average frequency of galaxies of the query class among the {\it Rank} galaxies returned by the algorithm.}
\label{spiral_in_elliptical_100}
\end{center}
\end{figure}

\begin{figure}[ht]
\begin{center}
\includegraphics[scale=0.55]{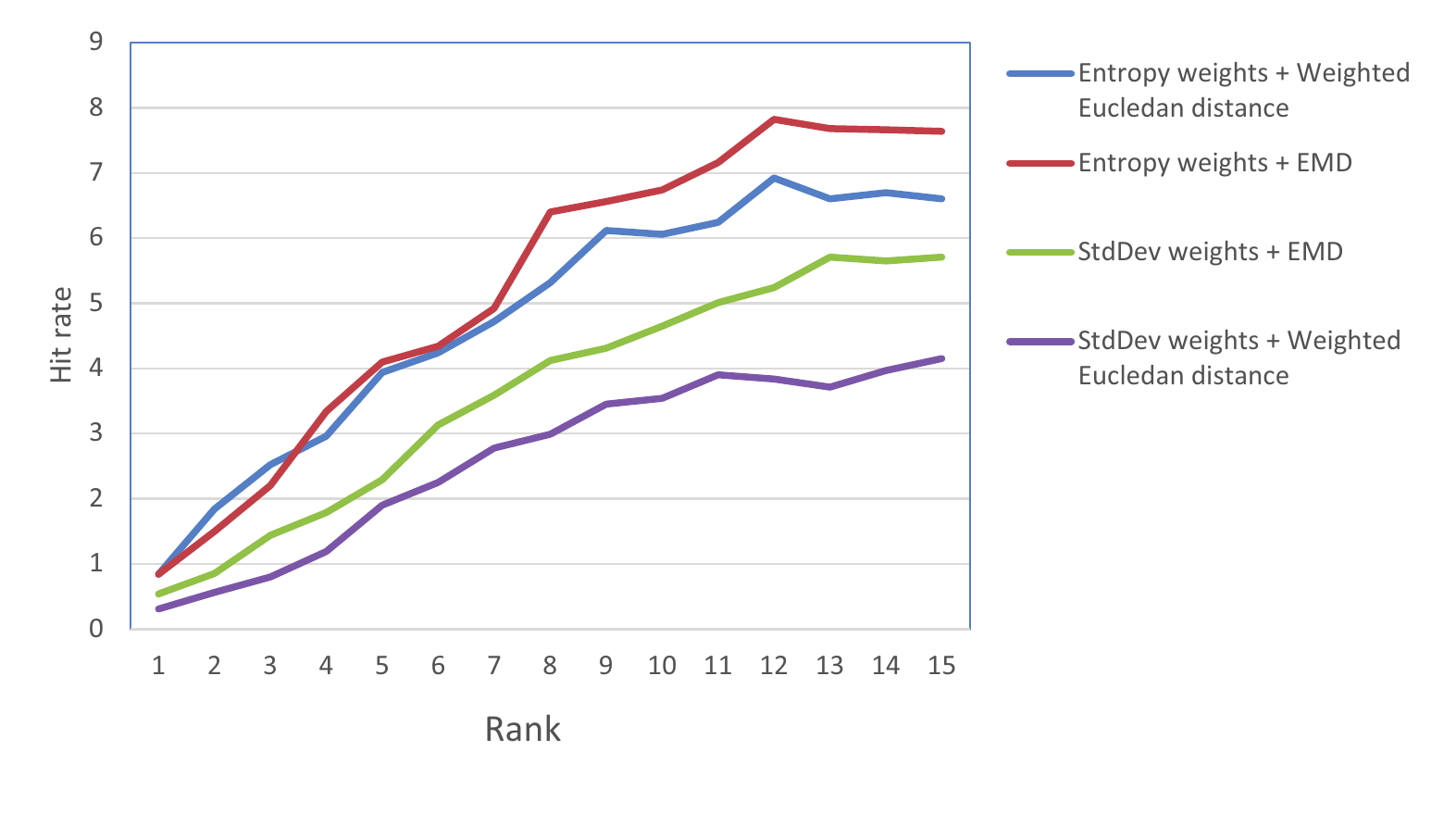}
\caption{Hit rate when using 100 spiral galaxies as ``regular'' galaxies and 10 randomly selected elliptical galaxies as ``peculiar'' galaxies. The hit rate is the average number of query galaxies among the {\it Rank} galaxies returned by the algorithm for each query galaxy.}
\label{elliptical_in_spiral_100}
\end{center}
\end{figure}

The small dataset of 100 spiral and 100 elliptical galaxies was used in two different ways; when the elliptical galaxies are considered ``regular'' and spiral are considered ``peculiar'', and then again when the spiral galaxies are considered ``regular'' and the elliptical galaxies are considered peculiar. In each of these datasets 10 ``peculiar'' galaxies were randomly combined with the 100 ``regular'' galaxies, and the experiment was repeated 100 times such that in each run a different galaxy was used as the query galaxy, and different 10 ``peculiar'' galaxies were randomly combined with the ``regular'' galaxies. That was repeated for the different ranks to check the performance and behavior of the algorithm. Figures~\ref{spiral_in_elliptical_100} and~\ref{elliptical_in_spiral_100} show the average number of returned galaxies that are morphologically similar to the query galaxy when the ``peculiar'' class is spiral galaxies, and when the peculiar class is elliptical galaxies, respectively.

As the figure shows, using earth movers distance (EMD) outperformed the Euclidean distance. When using entropy weights and EMD the algorithm returned an average of 0.72 spiral galaxies when the rank was 1 (meaning that the query returned just a single galaxy), and $\sim$4.4 spiral galaxies when the rank was 10. When the elliptical galaxies were considered the ``peculiar'' class, $\sim$6.74 galaxies of the 10 galaxies returned by the query (rank 10) where of the same morphological type as the query galaxy (elliptical).

\begin{figure}[ht]
\begin{center}
\includegraphics[scale=0.55]{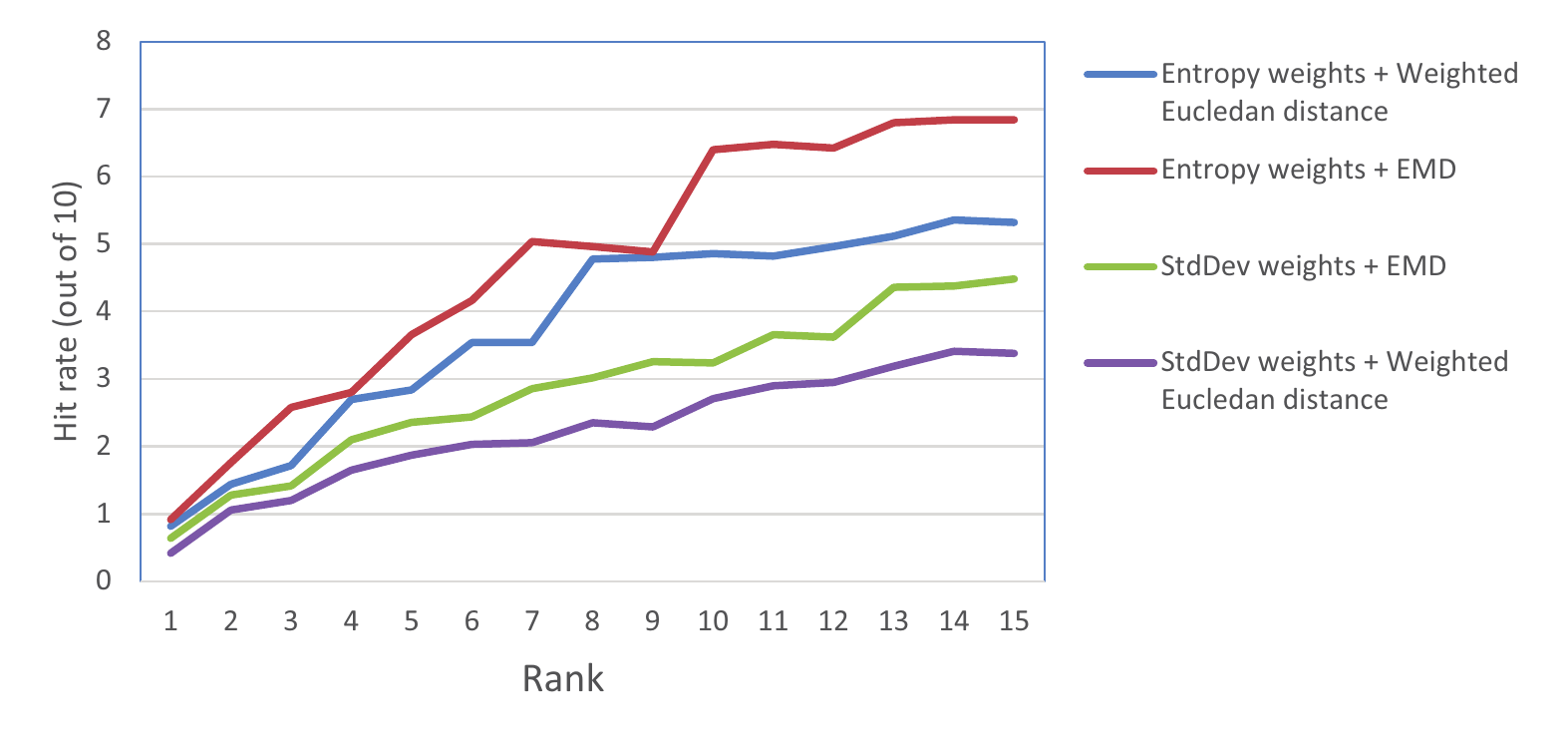}
\caption{Hit rate when using 100 elliptical galaxies as ``regular'' galaxies and 10 ring galaxies as ``peculiar'' galaxies}
\label{rings_in_ellipticals}
\end{center}
\end{figure}

\begin{figure}[ht]
\begin{center}
\includegraphics[scale=0.55]{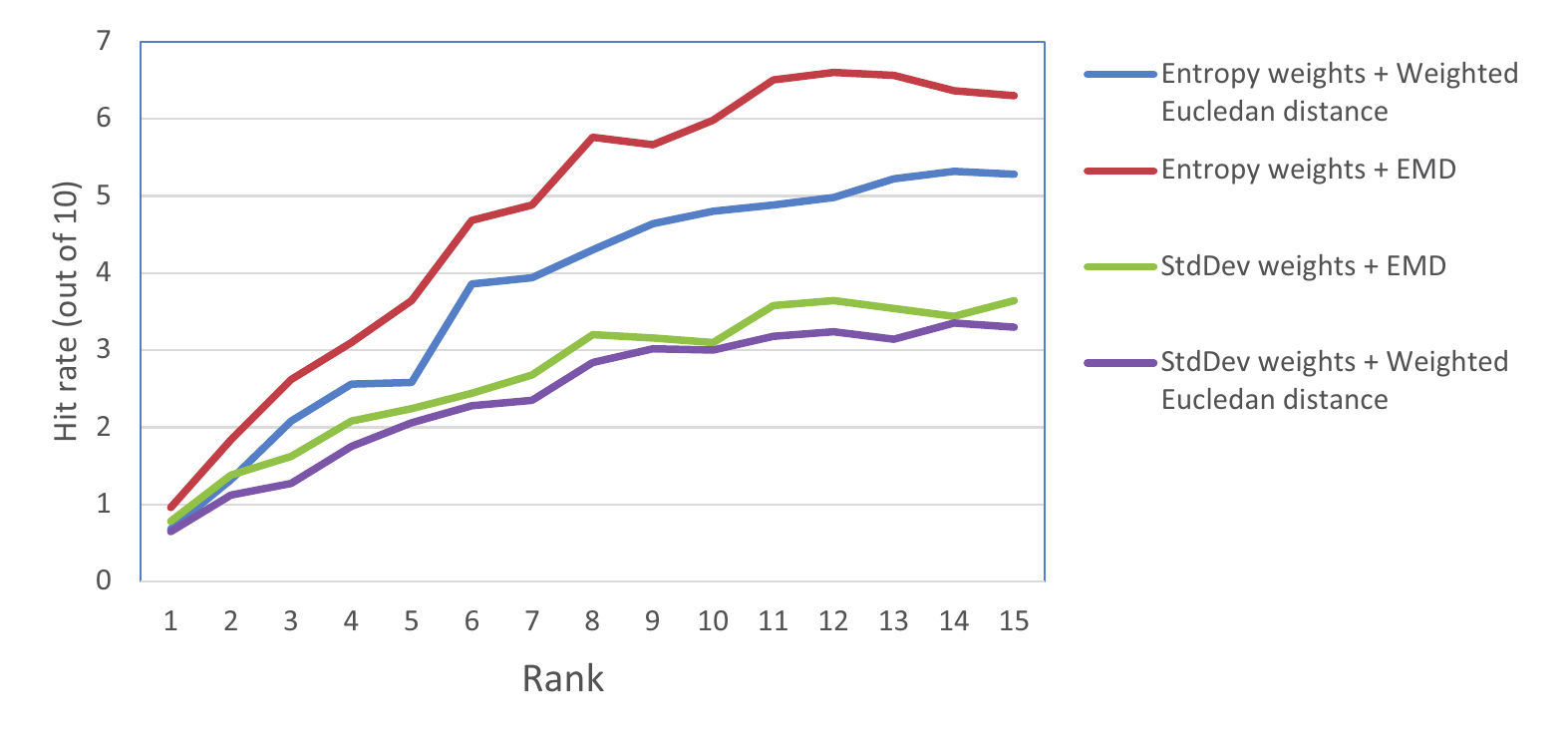}
\caption{Hit rate when using 100 spiral galaxies as ``regular'' galaxies and 10 ring galaxies as ``peculiar'' galaxies}
\label{rings_in_spiral}
\end{center}
\end{figure}

Figures~\ref{rings_in_ellipticals} and~\ref{rings_in_spiral} show the performance when the peculiar galaxies are ring galaxies, and the ``regular'' galaxies are elliptical and spiral galaxies, respectively. As before, 10 ring galaxies were combined with the 100 ``regular'' galaxies.

\begin{figure}[ht]
\begin{center}
\includegraphics[scale=0.55]{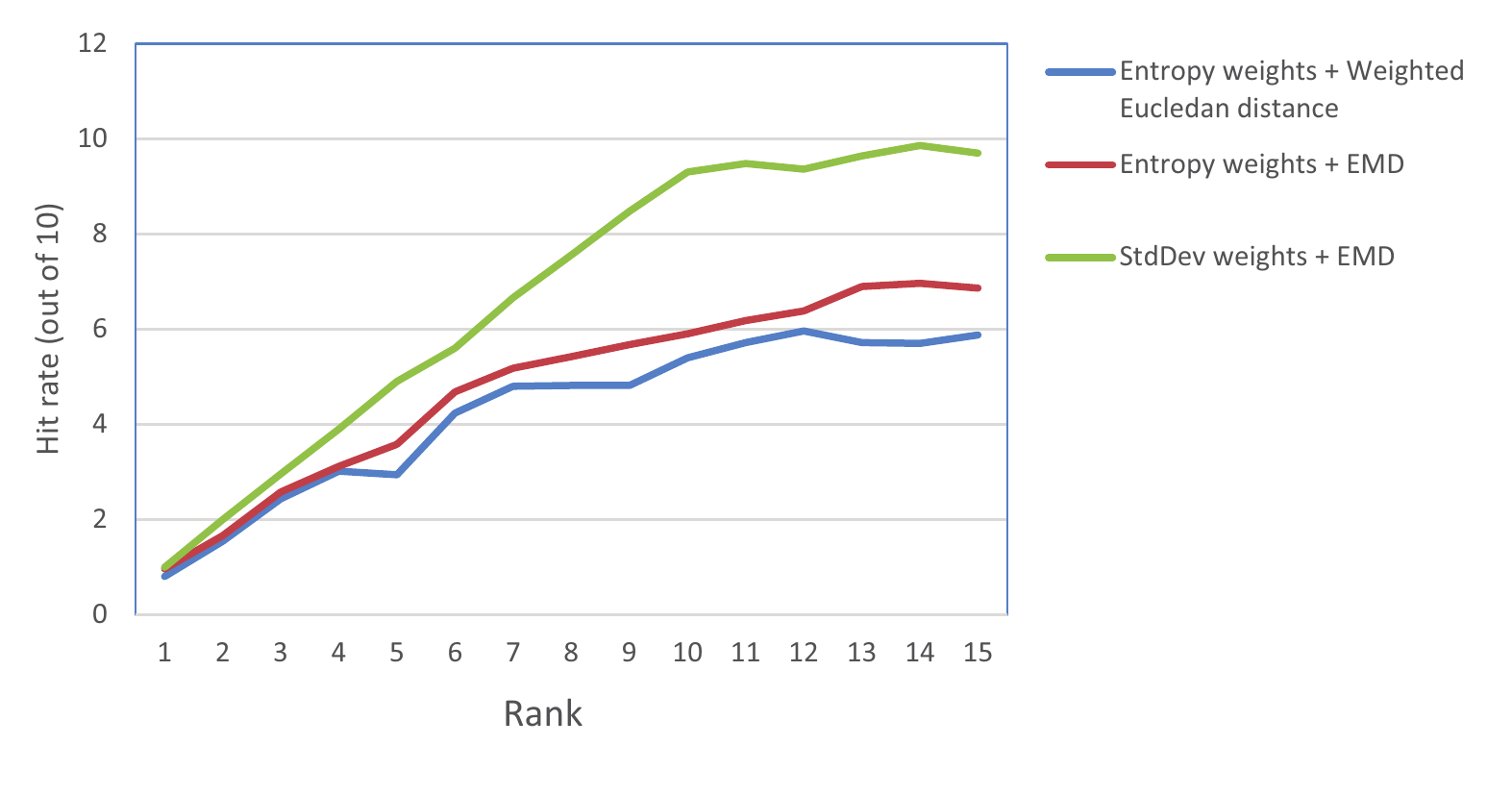}
\caption{Hit rate when using 100 elliptical galaxies as ``regular'' galaxies and 10 interacting galaxies as ``peculiar'' galaxies}
\label{mergers_in_ellipticals}
\end{center}
\end{figure}

\begin{figure}[ht]
\begin{center}
\includegraphics[scale=0.55]{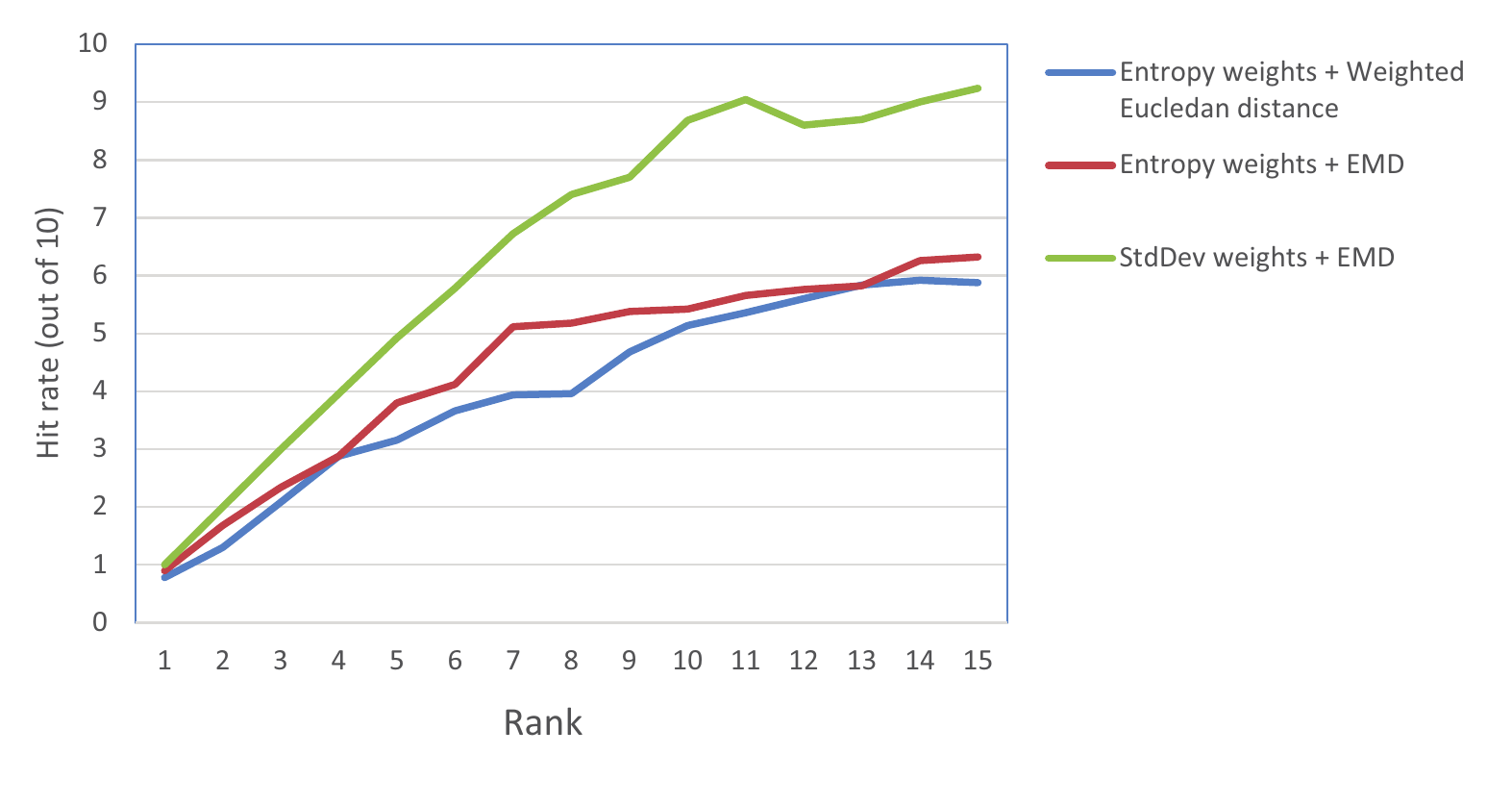}
\caption{Hit rate when using 100 spiral galaxies as ``regular'' galaxies and 10 interacting galaxies as ``peculiar'' galaxies}
\label{mergers_in_spiral}
\end{center}
\end{figure}

Figures~\ref{mergers_in_ellipticals} and~\ref{mergers_in_spiral} show the performance of the system when the ``peculiar'' galaxies are the interacting galaxies shown in Figure~\ref{mergers}, and the ``regular'' galaxies are 100 elliptical galaxies and 100 spiral galaxies, respectively.

\begin{figure}[ht]
\begin{center}
\includegraphics[scale=0.55]{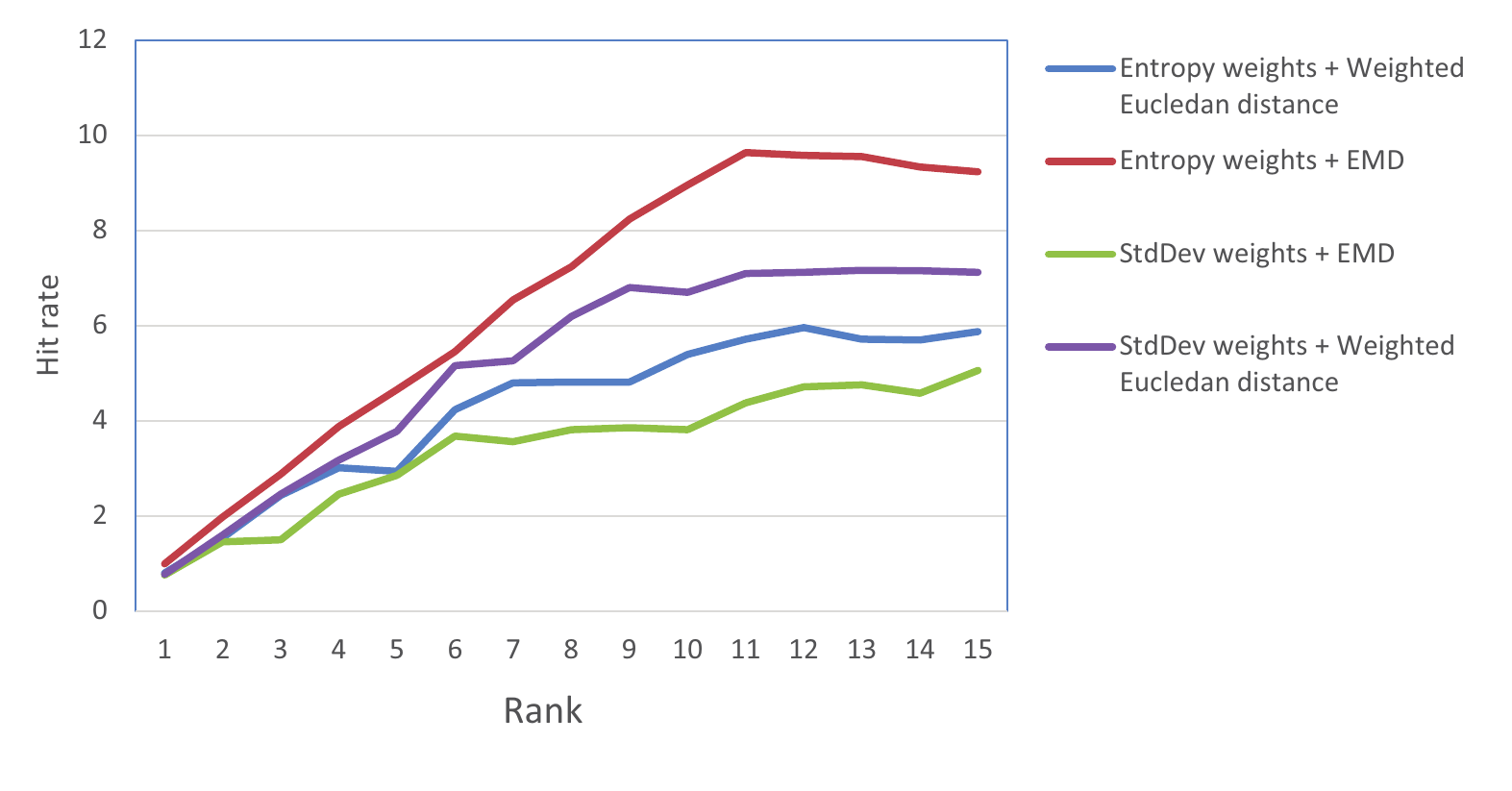}
\caption{Hit rate when using 200 elliptical and spiral galaxies as ``regular'' galaxies, and 10 interacting galaxies as ``peculiar'' galaxies}
\label{mergers_combined}
\end{center}
\end{figure}

\begin{figure}[ht]
\begin{center}
\includegraphics[scale=0.55]{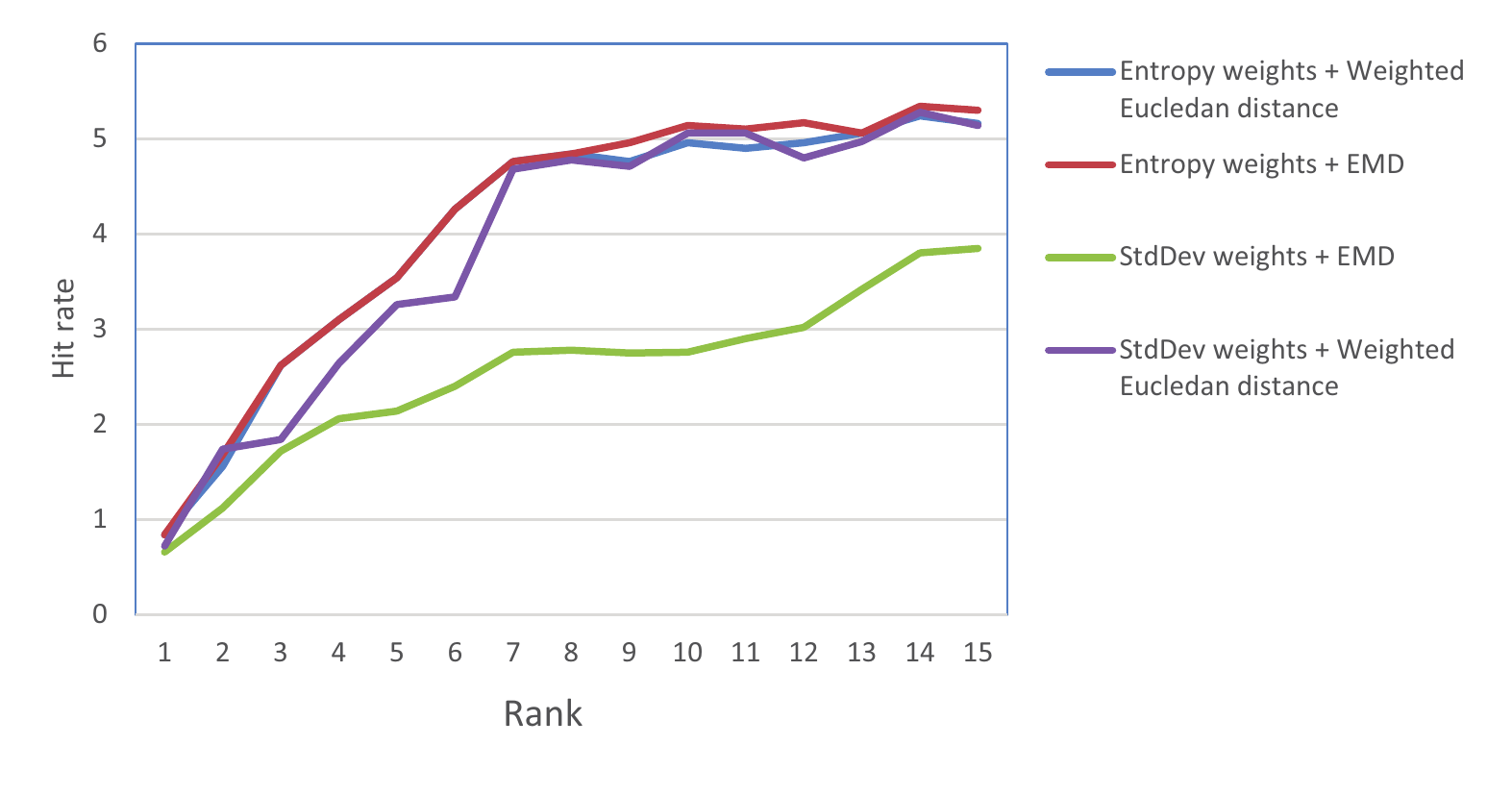}
\caption{Hit rate when using 200 elliptical and spiral galaxies as ``regular'' galaxies, and 10 ring galaxies as ``peculiar'' galaxies}
\label{rings_combined}
\end{center}
\end{figure}

In another experiment, the two sets of 100 spiral and 100 elliptical galaxies were combined into one dataset of 200 galaxies, and the experiment was repeated for the interacting and ring galaxies. The results of the experiments when the ``peculiar'' galaxies are ring galaxies or interacting galaxies are shown in Figures~\ref{mergers_combined} and~\ref{rings_combined}, respectively. As the figures show, the best performance was achieved when using the entropy weights, and measuring the distance using the Earth Mover Distance.

To test the performance of the algorithm in a larger set of galaxies, the ability of the algorithm to detect merging and ring galaxies was tested such that the peculiar galaxies were ring or merger galaxies, and three different sets of galaxies were used as the ``regular'' galaxies: a set of 4,000 spiral galaxies, a set of 4,000 elliptical galaxies, and a set of 10,000 objects identified as galaxies in SDSS DR8. 

The experiments were done using the entropy weights and EMD distances, such that 20 ``peculiar'' galaxies (galaxy mergers or ring galaxies) were used. Similarly to the other experiments, the 20 galaxies were used such that 10 galaxies were merged with the large set of ``regular'' galaxies, and one ``peculiar'' galaxy was used as the query galaxy. Each experiment was repeated 20 times such that in each run a different galaxy was used as the query galaxy.

\begin{figure}[ht]
\begin{center}
\includegraphics[scale=0.75]{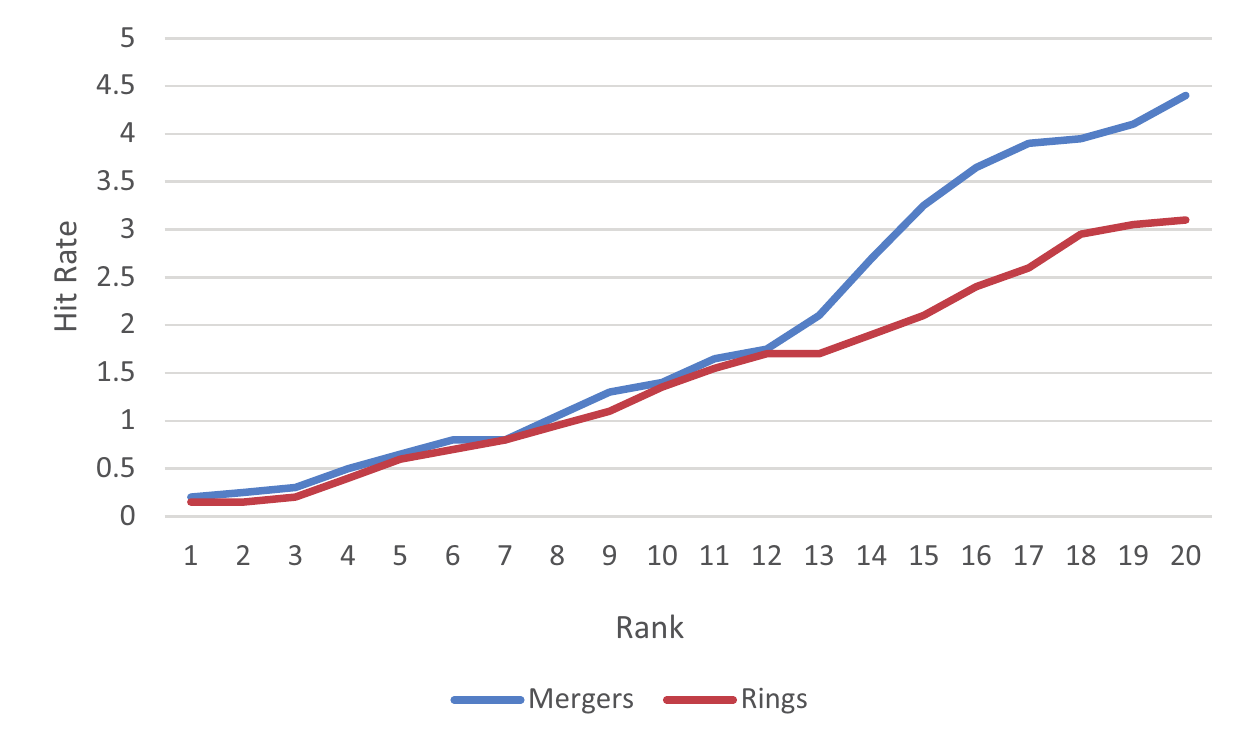}
\caption{Hit rate when using 4,000 elliptical galaxies as ``regular'' galaxies, and 20 ring or merging galaxies as ``peculiar'' galaxies}
\label{elliptical}
\end{center}
\end{figure}

\begin{figure}[ht]
\begin{center}
\includegraphics[scale=0.7]{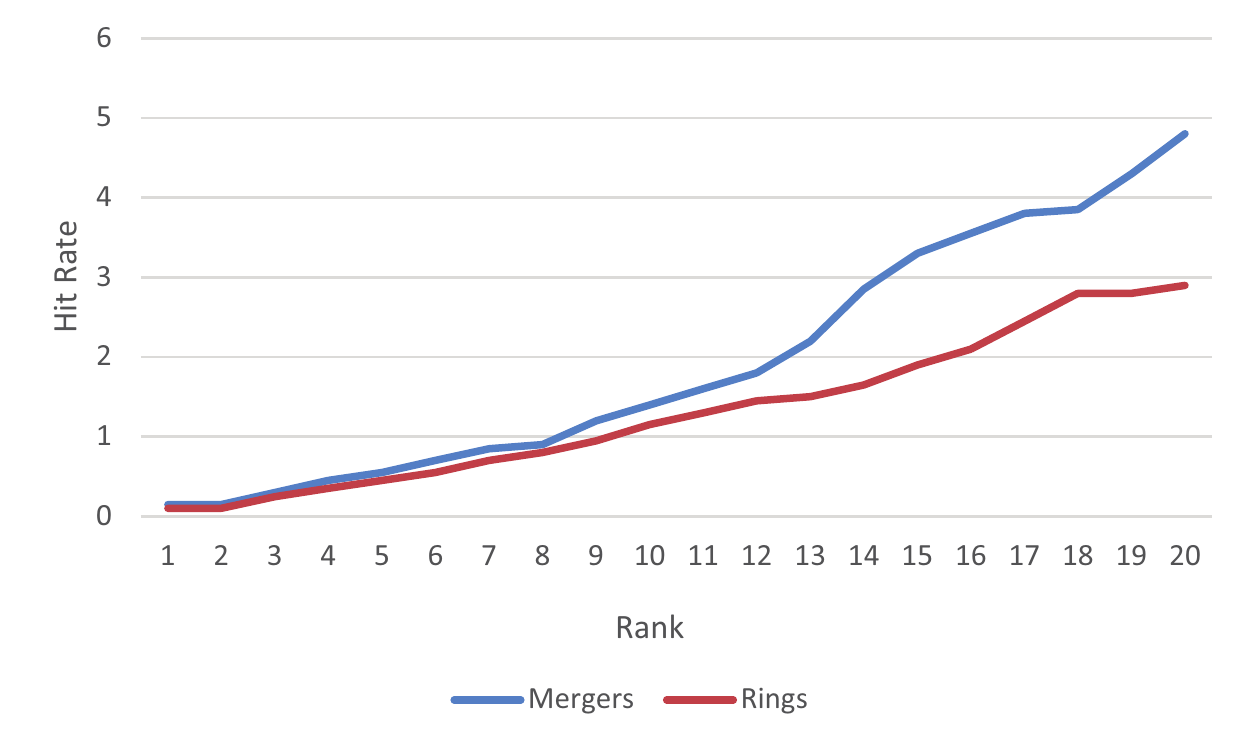}
\caption{Hit rate when using 4,000 SDSS spiral galaxies as ``regular'' galaxies, and 20 ring or merging galaxies as ``peculiar'' galaxies}
\label{spirals}
\end{center}
\end{figure}

Figures~\ref{elliptical} and~\ref{spirals} show the detection accuracy of ring and merging galaxies among datasets of 4,000 elliptical and 4,000 spiral galaxies, respectively. As the figures show, among a dataset of 4,000 spiral galaxies the algorithm was able to find $\sim$3 ring galaxies when the query galaxy was a ring galaxy, and $\sim$5 merging galaxies when the query galaxy was a galaxy merger.

\begin{figure}[ht]
\begin{center}
\includegraphics[scale=0.75]{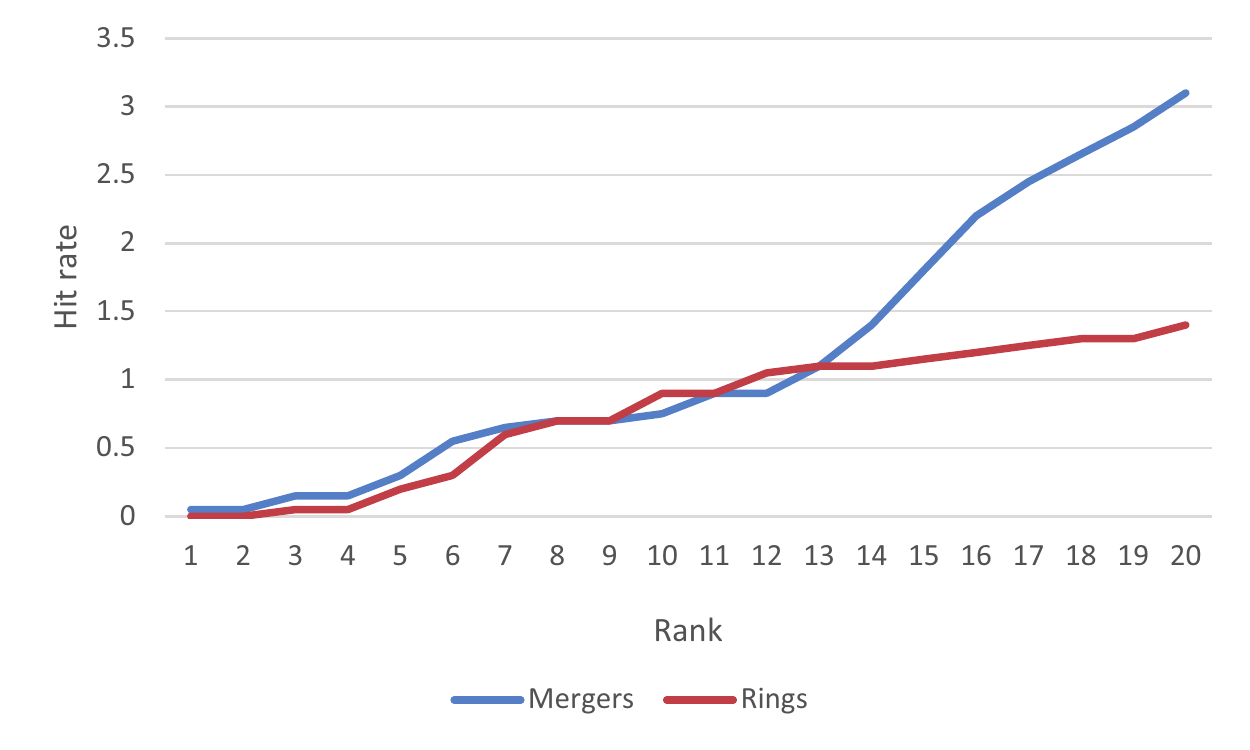}
\caption{Hit rate when using 10,000 SDSS galaxies as ``regular'' galaxies, and 20 ring or merging galaxies as ``peculiar'' galaxies}
\label{i_18}
\end{center}
\end{figure}

Figure~\ref{i_18} shows the detection accuracy of ring and merging galaxies among a dataset of 10,000 objects identified as galaxies by SDSS DR8, and have i magnitude of less than 18. As the figure shows, for both ring and merging galaxies, in most cases the algorithm was able to detect a galaxy similar to the query galaxy among the top 10 galaxies. For merging galaxies the algorithm was able to detect $\sim$3 galaxies similar to the query galaxy in the top 20 galaxies returned by the algorithm.

\begin{figure*}[ht]
\begin{center}
\includegraphics[scale=0.5]{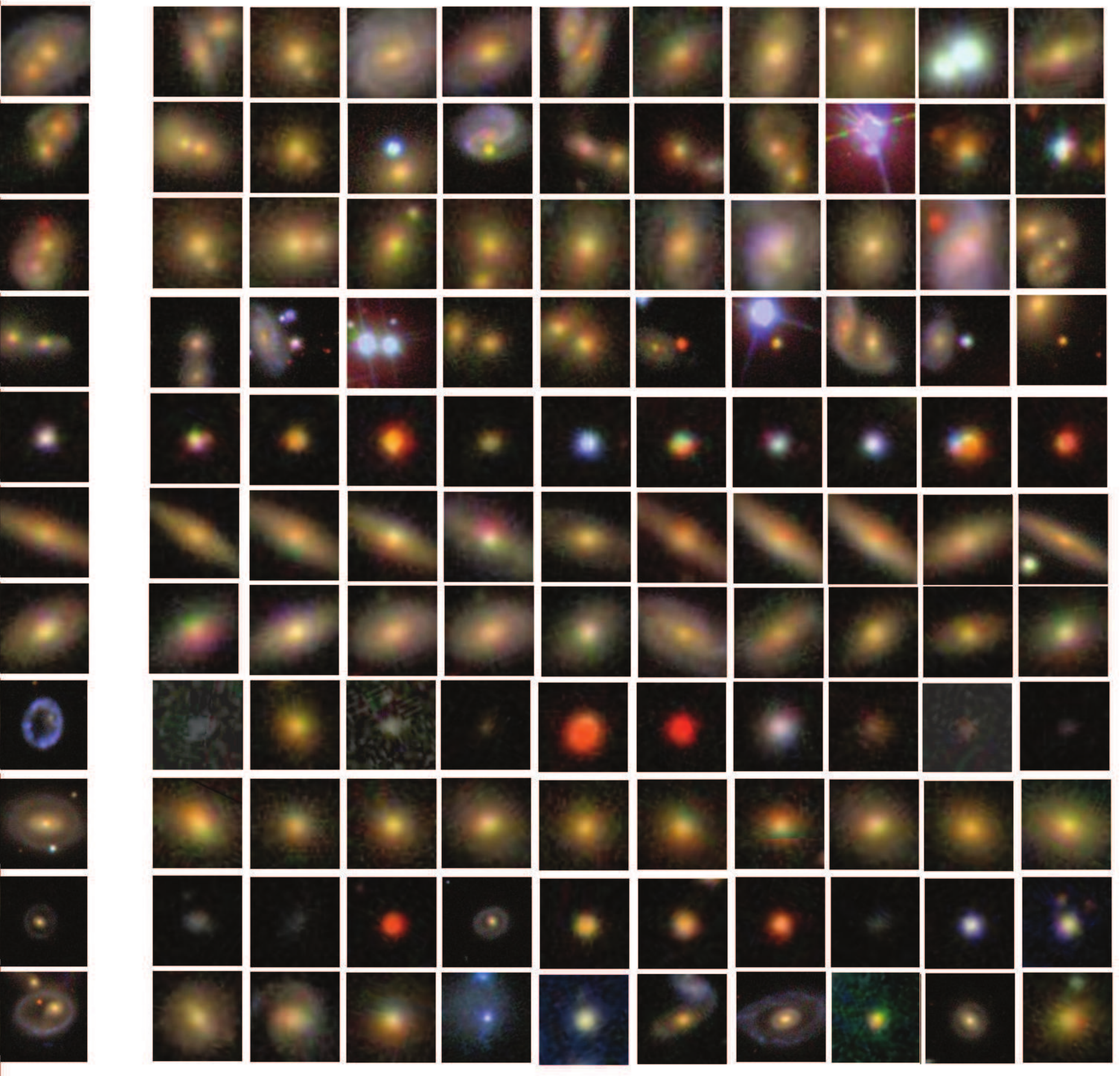}
\caption{Example of query results. The leftmost galaxy in each row is the query galaxy, and the other 10 galaxies in each row are the first 10 galaxies returned by the algorithm. The dataset is the dataset of 10,000 SDSS galaxies with i$<$18, combined with the ring and merger galaxies shown in Figures~\ref{ring_galaxies} and~\ref{mergers}.}
\label{examples}
\end{center}
\end{figure*}

Figure~\ref{examples} shows several query examples. The figure shows the galaxies returned by the query for several query galaxies. The dataset from which the galaxies were detected by the algorithm is the dataset of 10,000 SDSS galaxies used in the experiment shown inf Figure~\ref{i_18}, combined with the ring and merger galaxies shown inf Figures~\ref{ring_galaxies} and~\ref{mergers}.

The figure shows that in some cases the returned galaxies are not similar to the target galaxies. That is especially noticeable in the case of ring galaxies, which are not very common in the database. The images clearly show that noise has substantial effect on the performance of the algorithm, and many of the images returned by the algorithm are not necessarily similar to the query galaxy. Since analyzing image data is by nature a complex task for computing machines, and the analysis performed here is unsupervised, it is expected that noise will have substantial impact on the system, and the results returned by it will not be neither clean nor complete.

\subsection{Differences in size and luminosity}

Large databases of galaxy images are expected to be diverse, and to contain objects of different sizes and luminosities. An effective query should be able to return objects that are morphologically similar to the query object, regardless of their luminosity or size. To test the sensitivity of the system to the size and luminosity of the galaxies, the ring and merger images were modified such that the query image was changed while the other images were not changed. Each query was therefore performed such that the morphology of the galaxies was the same as the query galaxy, but the luminosity or size were different.

For luminosity, the query galaxy was modified such that the intensity of each pixel was reduced by 50\%. For size, the image was downscaled such that each side was reduced by 50\%, so that the size of the resulting image was 25\% of its original size. 

\begin{figure}[ht]
\begin{center}
\includegraphics[scale=0.65]{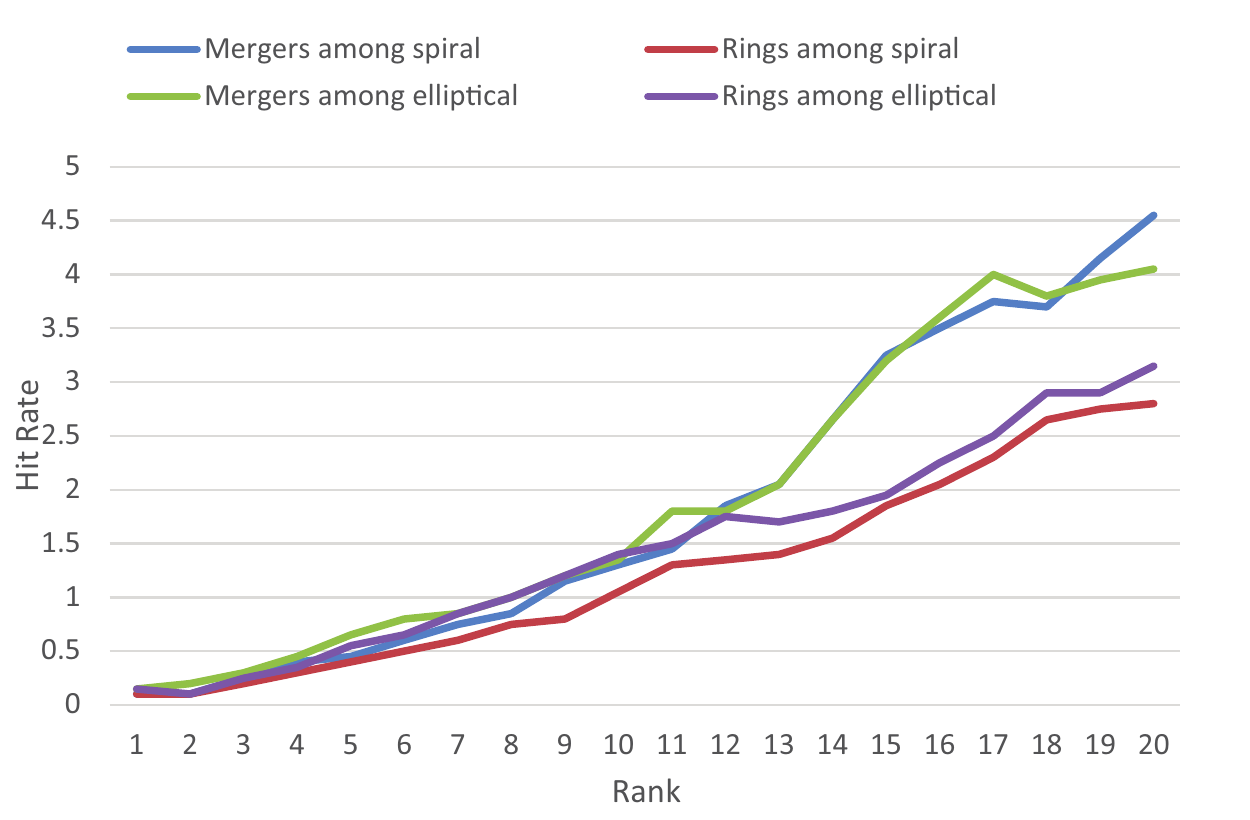}
\caption{Hit rate when using 4,000 elliptical galaxies or 4,000 spiral galaxies as ``regular'' galaxies, and 20 ring or interacting galaxies as ``peculiar'' galaxies. In each run the query galaxy image was made less bright, while the other galaxy images were not changed.}
\label{luminosity}
\end{center}
\end{figure}

\begin{figure}[ht]
\begin{center}
\includegraphics[scale=0.65]{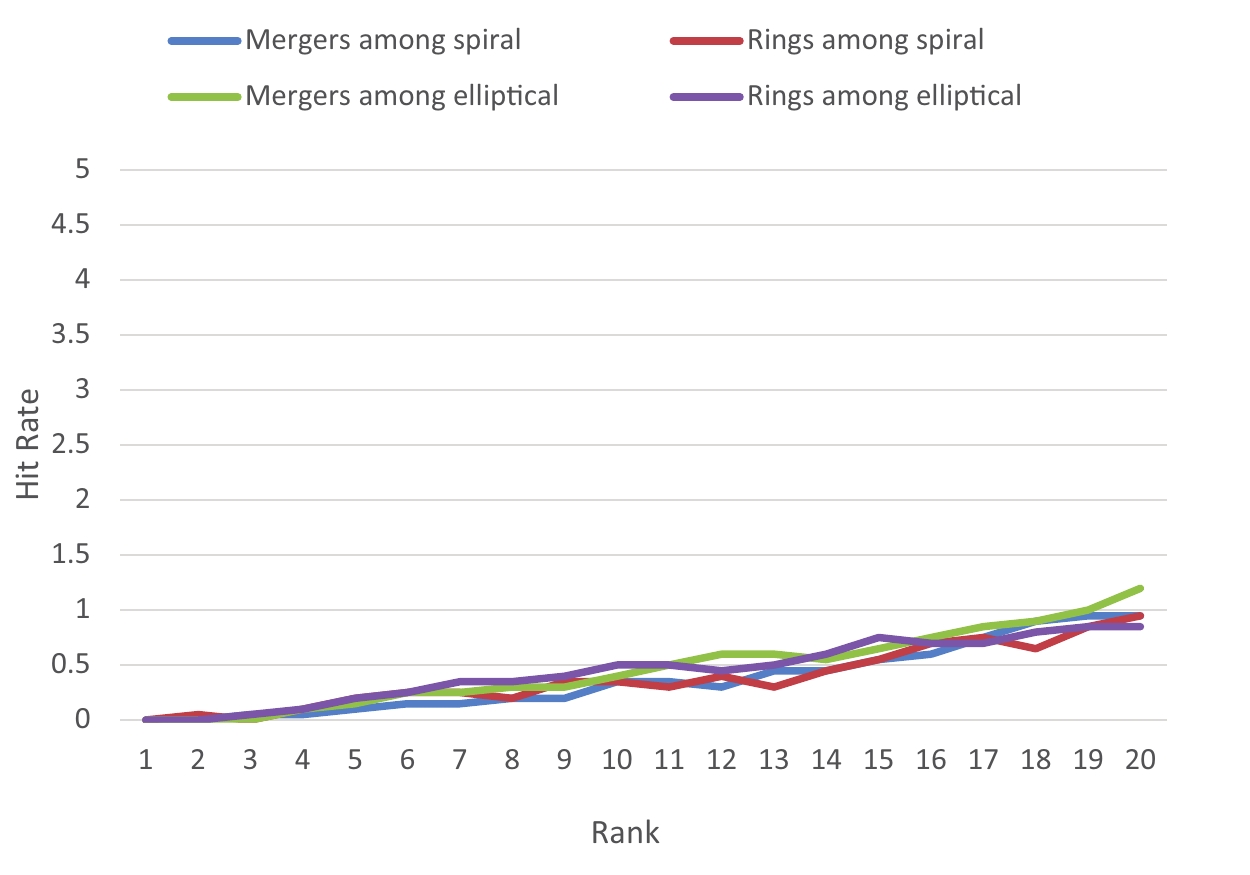}
\caption{Hit rate when using 4,000 elliptical galaxies or 4,000 spiral galaxies as ``regular'' galaxies, and 20 ring or interacting galaxies as ``peculiar'' galaxies. In each run the size of the query galaxy image was reduced, while the other galaxy images were not changed.}
\label{size}
\end{center}
\end{figure}

The performance of the algorithm when changing the brightness of the images are displayed by Figure~\ref{luminosity}, and Figure~\ref{size} shows the performance of the algorithm when the size was changed. As the figures show, changing the luminosity had a mild effect on the performance of the algorithm, showing that the algorithm is not sensitive to the brightness of the image. Reducing the size of the query images, on the other hand, had a strong negative effect on the performance.

The size of the images can clearly affect the efficacy of the algorithm. However, the size of the objects can be scaled to a certain consistent size. For instance, the images used in this study, taken from the catalog of elliptical and spiral galaxies \citep{kuminski2016computer}, were all downscaled to the size of 120$\times$120 pixels. Therefore, the galaxy images in the system need to be scaled to a certain consistent size as was done in \citep{kuminski2016computer}. Then, the query galaxy also needs to be scaled to the same size. If the spatial resolution does not allow scaling of the galaxy image to the standard size without artifacts, the performance of the algorithm will be affected.

As mentioned in Section~\ref{data}, the dimensionality of the galaxy images used in this experiment is 120$\times$120 pixels. The scaling can be also done automatically as was done in \citep{kuminski2016computer}, but all galaxy images in the system needs to be of a certain consistent size as was done in the experiment described in this paper or in \citep{shamir2014automatic,kuminski2016computer}.

Repeating the experiment described in Figures~\ref{spiral_in_elliptical_100} and~\ref{elliptical_in_spiral_100} such that the images were smoothed by a median filter with window size of 9$\times$9 provided similar results, and did not lead to an improvement.

\subsection{Completeness}

\begin{figure}[ht]
\begin{center}
\includegraphics[scale=0.65]{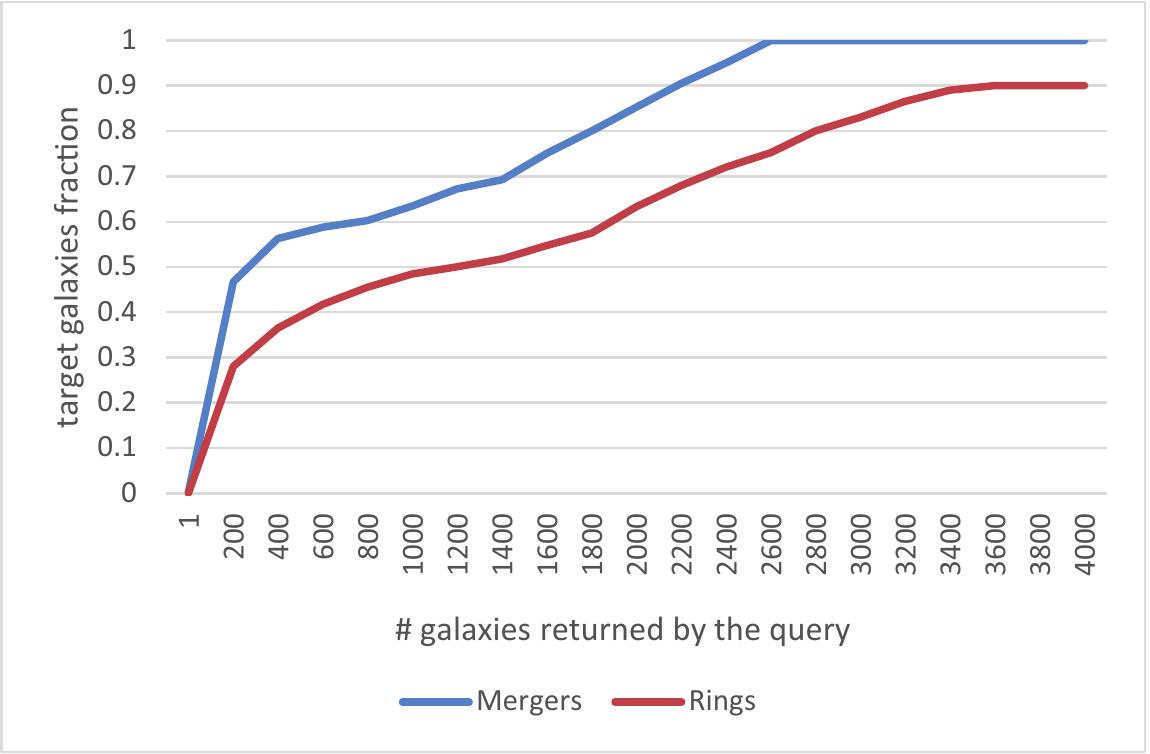}
\caption{Completeness when combining 20 interacting galaxies or 20 ring galaxies with 10,000 SDSS objects with i$<$18. The Y axis shows the number of the target galaxies (mergers or ring) returned by the algorithm divided by the total number of target galaxies.}
\label{completeness_i18}
\end{center}
\end{figure}

The goal of the algorithm is to reduce the database into a smaller subset in which the frequency of galaxies with morphology similar to the morphology of the query galaxy is substantially higher. However, the subset returned by the algorithm can be incomplete, leaving a certain number of galaxies with similar morphology outside of the subset returned by the algorithm. Naturally, it can be expected that when the subset returned by the algorithm is larger, it will include more of the target galaxies. The completeness can be measured simply by the average number of galaxies from the ``peculiar'' class returned by the query, divided by the total number of ``peculiar'' galaxies in the database. Figure~\ref{completeness_i18} shows the completeness of the list of galaxies returned by the algorithm when ring and merging galaxies are combined with a database of 10,000 SDSS galaxies.

As the figure shows, completeness is 100\% or close to it when the query returns $\sim$35\% of the data. These results show that achieving completeness with the algorithm is impractical, as in most digital sky surveys 35\% of the initial data is still far too large to allow practical manual analysis. On the other hand, when using the algorithm to reduce the dataset to 2\% of its initial size, it contained $\sim$50\% of the target galaxies when the query image was an image of an interacting galaxy, and $\sim$25\% of the target galaxies when the query image is an image of a ring galaxy.

\begin{figure}[ht]
\begin{center}
\includegraphics[scale=0.65]{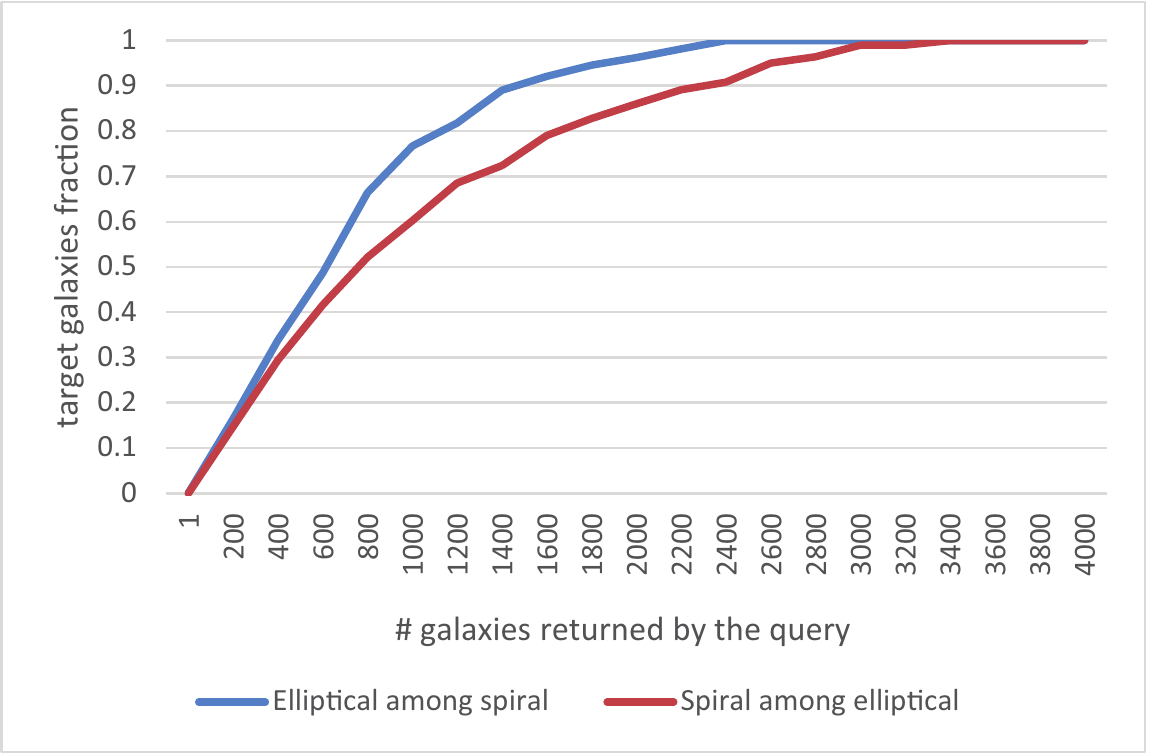}
\caption{Completeness when combining 1000 spiral galaxies with 4,000 elliptical galaxies, and when combining 1,000 elliptical galaxies with 4,000 spiral galaxies. The Y axis shows the number of the target galaxies (elliptical or spiral) returned by the algorithm divided by the total number of target galaxies.}
\label{completeness}
\end{center}
\end{figure}

To test the completeness on databases with more than 20 peculiar images, the completeness was also tested when using 4,000 spiral galaxies and ``regular'' galaxies and 1,000 elliptical galaxies as ``peculiar'', and vise versa. Figure~\ref{completeness} shows the completeness in that experiment. The graph shows that completion is achieved when the initial dataset is reduced to $\sim$2,500 galaxies (62.5\%) with the elliptical galaxies, and $\sim$3000 galaxies (75\%) when the target galaxies are spiral galaxies.

\subsection{Computational complexity and scalability}

The computational complexity of the system can be separated to the computational complexity of the numerical image content descriptors, which can be done off-line, and the computational complexity of the query itself, which needs to be done when the user submits a query.

Computing the numerical image content descriptors of a single 120$\times$120 galaxy image takes $\sim$45 seconds using a single Intel Core-i7 processor \citep{shamir2009automatic}. The numerical image content descriptors can be easily computed in parallel with negligible overhead \citep{Sha08,shamir2009automatic}, so the response time of computing the numerical image content descriptors is a function of the number of processors. A mid-size cluster of 320 processors used in this study can process over 600,000 galaxies per day, and analyzing millions of galaxies is clearly practical \citep{kuminski2016computer}. Computing the numerical image content descriptors can be done off-line, when the galaxy images are acquired are stored in the database, so that the response time of a user query is not dependent on the time required to compute the numerical image content descriptors.

The computational complexity of the query itself depends on computing the dissimilarity between the query galaxies to all galaxies in the database, as well as sorting the dissimilarity values to return a list of the most similar objects. Sorting the values can be done in complexity of $O(n \log n)$, where $n$ is the number of galaxies in the database. 

The complexity of computing the distances between the query galaxy and the database galaxies depends on the distance used for the dissimilarity measurement. If the Euclidean distance is used, the computational complexity of computing the dissimilarity between the query galaxy and any galaxy in the database is $O(m)$, where $m$ is the size of the feature vector. The computational complexity of computing the dissimilarities between the query galaxy and all galaxies in the database is therefore $O(n\cdot m)$.

The EMD distance can be computed in $O(m^2)$ operations \citep{ling2007efficient}, and therefore the process of computing the dissimilarities with all galaxies in the database is $O(n \cdot m^2)$. Although such low computational complexity should allow querying large databases of galaxies, computing the dissimilarities can also be parallelized with very low overhead when needed. The simple parallelization can be done such that each processor computes the dissimilarities to a different set of galaxies in the database, and when all dissimilarity values are computed they can be sorted to provide a list of the most similar galaxies.

\section{Conclusion}
\label{conclusion}

While digital sky surveys acquire very large image databases \citep{brescia2014dameware,longo2014data}, the natural difficulty of computing machines to analyze image data introduces an obstacle to researchers who wish to turn these data into scientific discoveries. As such databases can contain the images of very many galaxies, rare objects can be studied more effectively by identifying galaxies of similar morphology in the database. Therefore, once a galaxy of rare morphology of interest is identified, a researcher can use databases created by digital sky surveys to identify similar objects. However, identifying such objects among millions or even billions of other objects is impractical without automation.

Here we propose the task of automatic identification of galaxies that are visually similar to a given query galaxy of interest. Such system can assist researchers to identify galaxies similar to a certain galaxy of their interest, and therefore allow studying it with a population of galaxies. The ability of computers to identify galaxies in large databases based on a given query galaxy of interest can utilize the discovery power of digital sky surveys and improve their scientific return, enhancing existing and future data management systems by providing their user with stronger research capabilities.

Automatic identification of galaxies based on a user-provided query image that is not known to the system before the query is made requires unsupervised machine learning. Unsupervised machine learning is a harder problem than supervised learning \citep{bailey1995unsupervised} since the patterns need to be discovered without ground truth that can be used to separate samples and profile the different patterns. Also, working with non-structured data such as image data, as well as the complex nature of galaxy images, increase the difficulty of the problem further, and therefore it is expected that an algorithm aiming to solve that problem will neither clean nor complete, and noise will have substantial effect on its performance.

While the list of galaxies returned by the algorithm is not expected to be completely clean, the frequency of the galaxies of the morphology of interest is expected to be far higher than the frequency of their occurrence in the database. Preliminary results show that the algorithm can reduce the data substantially to return a much smaller dataset in which the frequency of the galaxy type of interest is high, allowing a practical second step of manual inspection to identify those galaxies. Such a system can be part of the data management of digital sky surveys, or can be implemented as an independent system by importing galaxy images from the digital sky survey and maintaining an independent large database of numerical image content descriptors. 

The software and source code of the method can be accessed at \url{http://vfacstaff.ltu.edu/lshamir/downloads/udat}.

\section{Acknowledgment}

This study was supported by NSF grant IIS-1546079.

\bibliographystyle{pasp}
\bibliography{ms}


\end{document}